\documentclass[conference]{IEEEtran}
\IEEEoverridecommandlockouts
\usepackage{amsmath,amssymb,amsfonts}
\usepackage{algorithmic}
\usepackage{graphicx}
\usepackage{textcomp}
\usepackage[nolist]{acronym}
\usepackage{xcolor}
\usepackage{hyperref}
\usepackage{cleveref}
\usepackage{stfloats}
\usepackage{todonotes}
\usepackage{pgffor}

\usepackage{booktabs}
\usepackage[sortcites, sorting=none, style=numeric-comp]{biblatex}

\usepackage[caption=false, font=scriptsize, captionskip=0pt]{subfig} 

\def\BibTeX{{\rm B\kern-.05em{\sc i\kern-.025em b}\kern-.08em
    T\kern-.1667em\lower.7ex\hbox{E}\kern-.125emX}}

\newcommand{\dsname}{RAIL-5G}

\title{RAIL-5G: A Challenging Real-World Dataset and Benchmark for AI-Based 5G Indoor Localization
\thanks{The authors would like to thank the Federal Ministry of Research, Technology, and Space (BMFTR) for supporting the xG‑RIC project as part of the research program Communication Systems “Souverän. Digital. Vernetzt.” (grant number 16KIS2429K) and the 5GILABB - 5G Industrial Lab Bischheim project (Grant No. 01MJ22017A).}
}

\def\authorrefmark#1{\ensuremath{^{\textbf{#1}}}}
\author{
    Jonas Pirkl\authorrefmark{1,}\authorrefmark{3},
    Andreas Eidloth\authorrefmark{1},
    Maximilian Kasparek\authorrefmark{1},
    Tobias Feigl\authorrefmark{1,}\authorrefmark{2},
    and Christopher Mutschler\authorrefmark{1,}\authorrefmark{3}\\
    {\tt\footnotesize\{firstname.lastname\}@iis.fraunhofer.de}\\ 
    \IEEEauthorblockA{\authorrefmark{1}
        Fraunhofer Institute for Integrated Circuits (IIS), 
        Division Positioning and Networks, 
        90411 Nürnberg, Germany
        }
    \IEEEauthorblockA{\authorrefmark{2}
        Programming Systems Lab,
        Friedrich-Alexander-Universität Erlangen-Nürnberg, 
        91052 Erlangen, Germany 
    }
    \IEEEauthorblockA{\authorrefmark{3}
        Department Computer Science \& Artificial Intelligence, Technische Universität Nürnberg, 90461 N\"urnberg, Germany
    }
}

\begin{acronym}[tdma]
    \acro{csi}[CSI]{channel state information}
    \acro{nn}[NN]{neural network}
    \acro{los}[LOS]{line-of-sight}
    \acro{nlos}[NLOS]{non-line-of-sight}
    \acro{pdr}[PDR]{pedestrian dead reckoning}
    \acro{cc}[CC]{channel charting}
    \acro{fp}[FP]{fingerprinting}
    \acro{mpc}[MPC]{multi path component}
    \acro{mae}[MAE]{mean average error}
    \acro{cir}[CIR]{channel impulse response}
    \acro{mse}[MSE]{mean squared error}
    \acro{adp}[ADP]{angle-delay profile}
    \acro{tdoa}[TDoA]{time-difference of arrival}
    \acro{toa}[ToA]{time of arrival}
    \acro{pca}[PCA]{principal component analysis}
    \acro{rssi}[RSSI]{Received Signal Strength Indicator}
    \acro{mlp}[MLP]{multi-layer perceptron}
    \acro{ml}[ML]{machine learning}
    \acro{adapos}[AdaPos]{Adaptive Positioning}
    \acro{sota}[SOTA]{state-of-the-art}
    \acro{bs}[BS]{base station}
    \acro{ru}[RU]{radio unit}
    \acro{ue}[UE]{user equipment}
    \acro{cnn}[CNN]{Convolutional Neural Network}
    \acro{usrp}[USRP]{Universal Software Radio Peripheral}
    \acro{pca}[PCA]{Principal Component Analysis}
    \acro{tsne}[t-SNE]{t-distributed stochastic neighbor embedding}
    \acro{ae}[AE]{autoencoder}
    \acro{umap}[UMAP]{uniform manifold approximation and projection}
    \acro{uageo}[UAGeo]{uncertainty-aware dimensionality reduction for channel charting with geodesic loss}
    \acro{srs}[SRS]{Sounding Reference Signal}
\end{acronym}

\begin{document}

\maketitle

\begin{abstract} 
Reliable radio-based indoor localization is essential for industrial automation, yet existing methods are often evaluated in environments that do not fully reflect the severe multipath and \ac{nlos} conditions encountered in real deployments.
This limits our understanding of how AI, i.e., modern geometric, unsupervised, and semi-supervised localization methods generalize to industrial 5G settings.
We present \dsname{}, a public real-world 5G NR \ac{cir} benchmark recorded in an operational railway maintenance facility.
\dsname{} covers an 811\,m$^2$ indoor area with eight distributed radio units, robot-mounted \ac{ue} trajectories, high-accuracy iGPS reference positions, and severe metal-induced multipath caused by train compartments and maintenance infrastructure.
We benchmark geometric \ac{tdoa}, supervised fingerprinting, unsupervised channel charting, and self-supervised wireless pretraining methods.
Our results highlight the difficulty of \dsname{}: on the full dataset, \ac{tdoa} reaches 7.02\,m \ac{mae}, the best unsupervised method reaches 2.46\,m, and supervised fingerprinting achieves 0.52\,m.
Self-supervised pretraining reduces the labeling effort by up to one order of magnitude, but the remaining gap to supervised learning shows that robust AI-based localization under severe multipath remains an open challenge. The dataset and code are available at \url{https://iis.fraunhofer.de/rail-5g}.
\end{abstract}


\acresetall

\vspace{-0.5cm}

\section{Introduction}

Accurate indoor localization is a key enabler for industrial automation, including asset tracking, mobile robotics, and logistics~\cite{Laoudias2018}.
5G NR is a promising platform for this task as it provides wide bandwidths, distributed infrastructure, and standardized reference signals.
However, industrial environments remain challenging: large metallic structures, machinery, vehicles, and narrow corridors create severe multipath and frequent \ac{nlos} conditions, which corrupt \ac{toa} estimates and degrade geometric methods such as \ac{tdoa}~\cite{Dardari2015}. Supervised fingerprinting can help but it requires expensive labeled data~\cite{fhg5g}.
Unsupervised and self-supervised methods aim to reduce this labeling effort: 
\Ac{cc} learns low-dimensional radio representations that preserve spatial neighborhood relations without reference positions~\cite{cc}, and recent extensions use Siamese or Triplet networks, dissimilarity metrics, temporal proximity, or auxiliary motion information~\cite{triplet-cc,adp-metric,pdr-cc}.
Self-supervised pretraining~\cite{sparstran} promises label-efficient localization returning top performance.
Despite this progress, it remains unclear how well these methods generalize to harsh industrial 5G environments, as existing public datasets often cover only parts of the problem, such as simulation, WiFi, massive-MIMO channel sounding, small-scale 5G measurements, or less severe propagation conditions.

\begin{figure}[t]
  \centering
  \includegraphics[trim=0 80 0 10, clip, width=\linewidth]{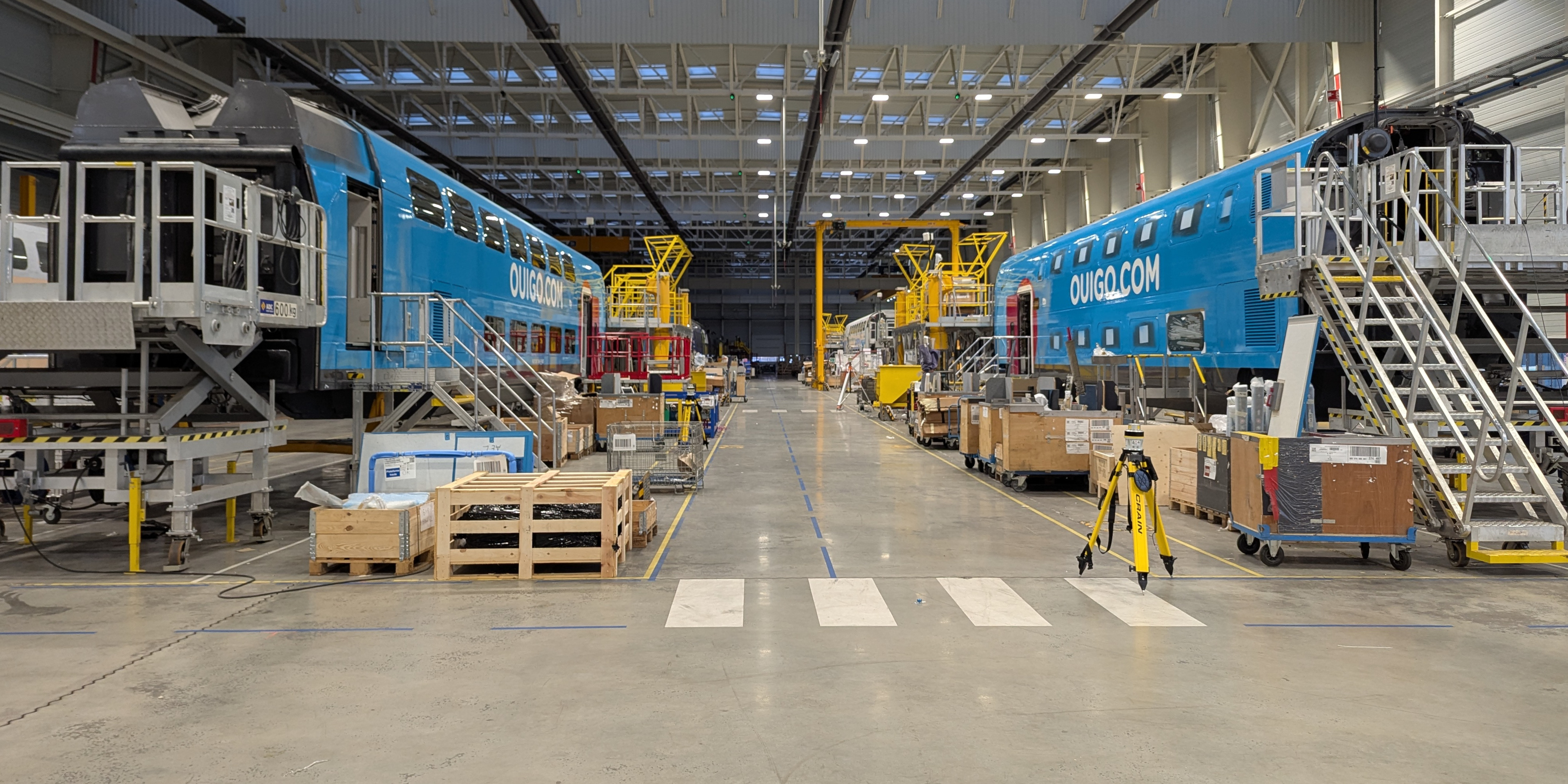}
  \vspace{-0.6cm}
  \caption{\dsname{} environment, viewed from RU~1 toward RU~4/5. Large metallic train compartments cause severe multipath propagation.}
  \label{fig:env-img}
  \vspace{-0.6cm}
\end{figure}

We propose \dsname{}, a public real-world 5G NR \ac{cir} benchmark recorded in an operational railway maintenance facility.
The dataset covers 811\,m$^2$ and contains uplink \ac{srs} measurements from a robot-mounted \ac{ue}, recorded by eight distributed \acp{ru}.
High-accuracy iGPS reference positions are provided at 40\,Hz.
The environment contains open areas and narrow corridors between metallic train compartments, resulting in heterogeneous \ac{los}/\ac{nlos} conditions and severe metal-induced multipath.
\dsname{} is a reality check for modern AI-based 5G indoor localization.
We compare it to existing public \ac{csi}/\ac{cir} datasets and benchmark geometric \ac{tdoa}, supervised fingerprinting, unsupervised channel charting, and self-supervised wireless pretraining methods.
Our results show that \dsname{} is challenging: on the full dataset, \ac{tdoa} reaches 7.02\,m \ac{mae}, the best unsupervised method reaches 2.46\,m, while supervised fingerprinting achieves 0.52\,m.
This gap shows that the \acp{cir} contain strong spatial information, but that current unsupervised methods struggle to exploit it under severe industrial multipath.
Self-supervised pretraining lowers the required number of labeled samples by up to 10$\times$, but does not fully close the gap to supervised learning yet.

The paper is structured as follows.
Sec.~\ref{sec:rel-work} reviews related datasets and benchmarks.
Sec.~\ref{sec:dataset} introduces the dataset.
Sec.~\ref{sec:results} benchmarks. Sec.~\ref{sec:conclusion} concludes.

\section{Related Work}\label{sec:rel-work}

\subsection{Simulated CSI Benchmarks}
Public datasets for radio-based localization are broadly grouped into simulated channel datasets and real-world \ac{csi}/\ac{cir} measurement campaigns.
Simulated datasets such as DeepMIMO~\cite{alkhateeb2019deepmimo}, MOCSID~\cite{makhlouf2025multi}, and xG-Dataset~\cite{conti-3gpp-compliant-2024} provide controllable propagation scenarios, accurate ground truth, and large amounts of labeled data.
So, they are valuable for reproducible benchmarking and method development.
However, simulated datasets do not fully capture hardware impairments, synchronization effects, and the irregular scattering behavior of real industrial environments with large metallic structures.

\subsection{Real-World CSI Benchmarks}
Real-world \ac{csi} datasets address some of these limitations.
OpenCSI~\cite{gassner2021opencsi} provides LTE \ac{csi} for indoor fingerprinting, while the Ultra-dense MaMIMO dataset~\cite{debast2020mamimo,debast2021ultradense} and related massive-MIMO measurements~\cite{mamimo} support supervised \ac{csi}-based positioning in controlled indoor environments.
DICHASUS~\cite{dichasus2021} provides a broad collection of real-world massive-MIMO \ac{csi} measurements and is an important benchmark for channel charting.
ESPARGOS~\cite{espargos2024} provides phase-coherent WiFi \ac{csi} datasets with synchronized antenna arrays and includes challenging setups with metal blockage and robot trajectories.
CAEZ-WiFi~\cite{caez-wifi2024} demonstrates distributed WiFi \ac{csi} acquisition for neural positioning and channel charting.
These datasets are valuable for \ac{csi}-based localization research, but they do not jointly provide real-world 5G NR \acp{cir}, large-scale industrial railway infrastructure, and severe multipath.

\subsection{5G Localization Datasets}
Recent 5G datasets are closer to industrial deployment scenarios.
CAEZ-5G~\cite{caez-5g2025} provides real 5G NR \ac{csi} measurements with commercial off-the-shelf user equipment and distributed O-RUs, enabling neural positioning, channel charting, and device classification.
However, its indoor scenario is a small lab/office area and does not capture large-scale industrial multipath.
The Fraunhofer 5G dataset~\cite{fhg5g} is the closest related dataset, as it provides real 5G NR \acp{cir} from an industrial warehouse-like environment with distributed receiving anchors and heterogeneous propagation.
However, the environment has only limited reflecting surfaces and wide open spaces.\\

Instead, \dsname{} targets a different and complementary setting:
It provides real-world 5G NR \acp{cir} from an operational railway maintenance facility with 8 distributed \acp{ru}, robot-mounted \ac{ue} trajectories, high-accuracy iGPS reference positions, and severe metal-induced multipath caused by train compartments and maintenance infrastructure.
By evaluating geometric \ac{tdoa}, supervised fingerprinting, unsupervised channel charting, and self-supervised pretraining on \dsname{}, it serves as a challenging benchmark for realistic AI-based 5G indoor localization.

\begin{figure}[t]
    \centering
    \includegraphics[width=\linewidth]{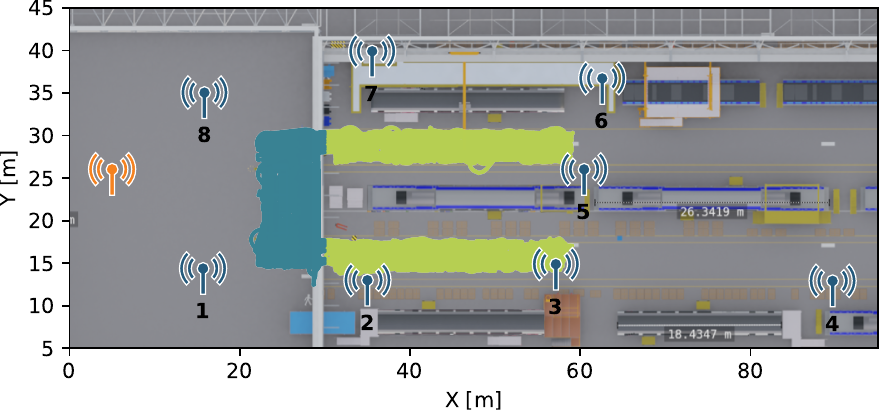}
    \vspace{-0.7cm}
    \caption{Map of the recording area. The environment consists of an open area on the left and two narrow corridors between metallic train compartments on the right. The distributed \acp{ru} are shown in dark blue.}
    \label{fig:env-map}
    \vspace{-0.4cm}
\end{figure}

\section{The \dsname{} Dataset}\label{sec:dataset}

This Sec. describes the recording environment (Sec.~\ref{sec:dataset:env}), the 5G measurement setup (Sec.~\ref{sec:dataset:setup}), the reference system (Sec.~\ref{sec:dataset:references}), the recorded trajectories (Sec.~\ref{sec:dataset:recordings}), and the dataset format (Sec.~\ref{sec:dataset:format}) of \dsname{}.

\subsection{Recording Environment}\label{sec:dataset:env}
\dsname{} was recorded at the SNCF Technicentre in Bischheim, France, a railway maintenance facility used for the maintenance and modernization of TGV trainsets.
The recording area spans approximately 38\,m $\times$ 22\,m, with 811\,m$^2$ covered by valid reference trajectories.
As shown in Fig.~\ref{fig:env-map}, the area consists of two distinct propagation regions.
The left side of the environment is an open area with \ac{los} to several \acp{ru}.
In contrast, the right side contains two narrow corridors between large metallic train compartments and maintenance infrastructure.
These corridors create strong reflections, frequent \ac{nlos} conditions, and dense multipath propagation.
An example corridor is shown in Fig.~\ref{fig:env-img}.
This combination of open space, metallic corridors, and heterogeneous \ac{los}/\ac{nlos} conditions makes the dataset particularly challenging for radio-based indoor localization in  real-world applications.

\subsection{5G Measurement Setup}\label{sec:dataset:setup}
The dataset was recorded using an Amarisoft 5G NR cell operating in band n77 at a carrier frequency of 3.95\,GHz.
The \ac{srs} was configured with 100\,MHz bandwidth and transmitted at a rate of 100\,Hz.
The uplink signals were captured using Fraunhofer's Mobile Testbed, a modular recording platform that can be attached to existing 5G systems to acquire positioning-relevant radio measurements.
The testbed was equipped with eight Benetel RAN550 n77u \acp{ru}, which were spatially distributed across the environment as shown in Fig.~\ref{fig:env-map}.
Each \ac{ru} received the uplink \ac{srs} signal and recorded the corresponding channel measurements at a sampling frequency of 122.88\,MHz.
The resulting setup mimics a distributed-SIMO uplink measurement system: a single mobile \ac{ue} transmits, while eight spatially distributed receivers observe the channel.
A Quectel RM502Q-AE 5G modem served as the \ac{ue}.
It was mounted on a mobile robot at approximately 0.42\,m above the floor.
During the recordings, the robot traversed the environment at an approximately constant speed of 1\,m/s.

\subsection{Reference System}\label{sec:dataset:references}
Ground-truth positions were recorded using a 7D Kinematic Metrology iGPS system.
The system uses a network of infrared laser transmitters and sensors to estimate positions through triangulation.
8 laser transmitters were deployed to cover the recording area.
The reference system provides sub-millimeter positioning accuracy at a sampling rate of 40\,Hz.
The reference positions are temporally aligned with the 5G measurements during preprocessing, as described in \Cref{sec:dataset:preprocessing}.

\begin{table}[t]
\centering
\caption{Overview of the recorded takes.}
\vspace{-0.25cm}
\begin{tabular}{llr}
\multicolumn{1}{l}{Take} & Description                       & \multicolumn{1}{l}{Data Points} \\ \midrule
\#1                        & Meander lower corridor            & 52,588                          \\
\#2                        & Meander free space                & 51,552                               \\
\#3                        & Meander upper corridor            & 54,963                               \\
\#4                        & Serpentine pattern lower corridor & 43,130                                \\
\#5                        & Serpentine pattern free space     & 25,809                                \\
\#6                        & Serpentine pattern upper corridor & 31,682                                \\
\#7                        & Random full area                  & 87,585                              \\
\#8                        & Coarse meander full area          & 68,045                            \\ \midrule
\multicolumn{2}{l}{Total} & 415,354
\end{tabular}
\vspace{-0.5cm}
\label{tab:takes}
\end{table}

\subsection{Recordings and Trajectories}\label{sec:dataset:recordings}
\begin{figure}[!tp]
    \centering
    \captionsetup[subfloat]{justification=centering, captionskip=2pt}
    \hfill \subfloat[Take \#1]{\includegraphics[width=0.45\linewidth]{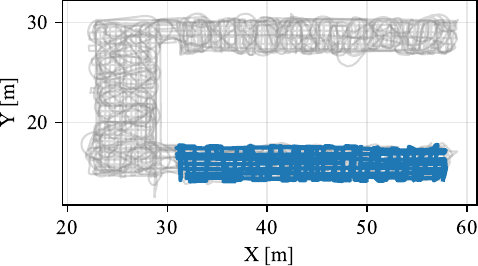}}\hfill
    \hfill\subfloat[Take \#2]{\includegraphics[width=0.45\linewidth]{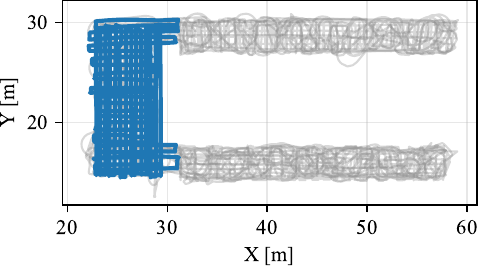}}\hfill \\ [-5pt]
    \hfill\subfloat[Take \#3]{\includegraphics[width=0.45\linewidth]{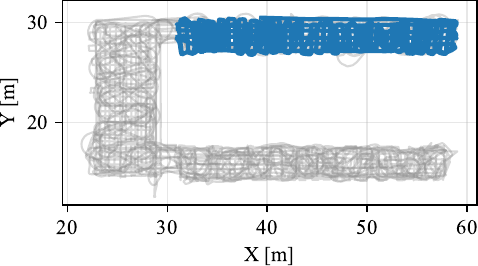}}\hfill
    \hfill\subfloat[Take \#4]{\includegraphics[width=0.45\linewidth]{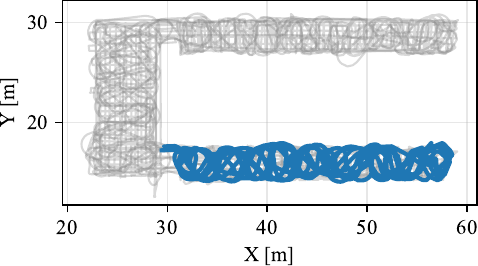}}\hfill \\[-5pt]
    \hfill\subfloat[Take \#5]{\includegraphics[width=0.45\linewidth]{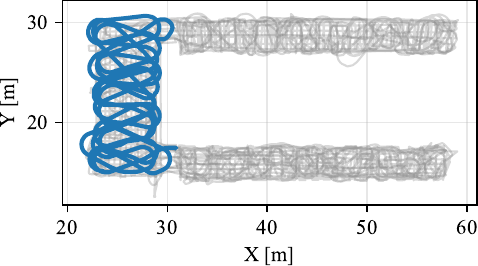}}\hfill
    \hfill\subfloat[Take \#6]{\includegraphics[width=0.45\linewidth]{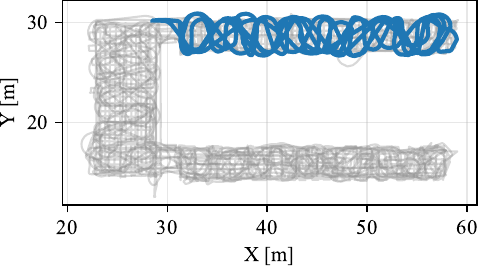}}\hfill\\[-5pt]
    \hfill\subfloat[Take \#7]{\includegraphics[width=0.45\linewidth]{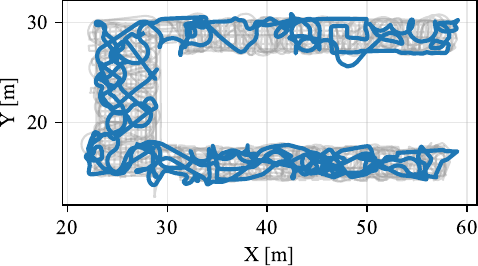}}\hfill
    \hfill\subfloat[Take \#8]{\includegraphics[width=0.45\linewidth]{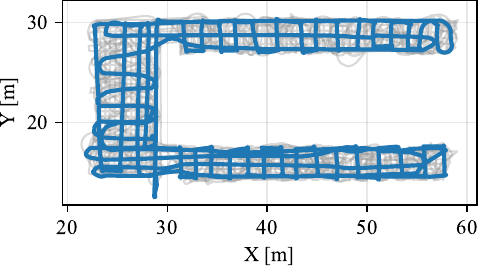}}
    \vspace{-0.1cm}
    \caption{Overview of full (grey) and subset (blue) data.}
    \label{fig:takes}
    \vspace{-0.5cm}
\end{figure}

We recorded 8 individual takes, summarized in Tab.~\ref{tab:takes}.
Takes \#1--\#6 cover individual subareas, while takes \#7 and \#8 cover the full recording area.
The trajectories include meander, serpentine, random, and coarse full-area motion patterns, providing both locally dense coverage and larger-scale traversal of the environment.
Fig.~\ref{fig:takes} shows an overview of the recorded takes.

\subsection{Preprocessing}\label{sec:dataset:preprocessing}
The raw frequency-domain channel measurements are transformed into the time domain to obtain \acp{cir}.
For each \ac{cir}, we estimate the \ac{toa} using the inflection point method~\cite{hofmann2009device} and align the \ac{cir} window such that the estimated direct path is placed at delay tap~10.
The aligned \acp{cir} are truncated to 96 delay taps, which captures the most relevant multipath components while keeping the data compact.
The \acp{cir} are recorded at 100\,Hz, while the iGPS reference system provides positions at 40\,Hz.
We linearly interpolate the reference positions to match the \ac{cir} timestamps.
As the reference system occasionally loses coverage for short intervals, the time difference to the preceding and following true reference measurement is stored for each sample.
This allows users to filter samples with excessive interpolation.
As the robot moves at approximately 1\,m/s, short interpolation gaps are generally acceptable; however, we recommend discarding samples where the total interpolation span exceeds 1\,s.
Due to the dense multipath environment, the \ac{toa} estimator may occasionally select a reflection or noise peak instead of the direct path.
This leads to invalid \ac{toa} estimates and misaligned \ac{cir} windows.
We apply sanity checks and peak-to-noise filtering to remove the majority of such cases, but a small number of erroneous \acp{cir} may remain.

\subsection{Dataset Format}\label{sec:dataset:format}
The dataset contains complex \acp{cir} for eight \acp{ru}, represented by real and imaginary components with 96 delay taps per receiver.
For each data point, the dataset further provides per-\ac{ru} \ac{toa} estimates, an antenna mask, a timestamp, an interpolated 2D reference position, and the interpolation time differences described in \Cref{sec:dataset:preprocessing}.
The \ac{toa} values are expressed relative to the earliest arriving signal across all valid \acp{ru}.
The antenna mask indicates whether the corresponding \ac{cir} and \ac{toa} are valid or missing due to signal blockage or estimation failure.
The \acp{cir} are not normalized and are therefore proportional to the received signal strength.
However, as the \ac{ue} may autonomously adjust its transmit power, amplitude comparisons across time should be used with caution.
Relative amplitude differences within a single sample, for example across \acp{ru}, remain meaningful.

\begin{table*}[t]
    \centering
    \caption{Localization results on 3 dataset splits (* are non-parametric methods and also fitted on the test set).}
    \vspace{-0.3cm}
    \resizebox{\textwidth}{!}{
    \begin{tabular}{lrrrrrr|rrrrrr|rrrrrr}
        & \multicolumn{6}{c|}{\textbf{Full}}
        & \multicolumn{6}{c|}{\textbf{Lower Corridor}}
        & \multicolumn{6}{c}{\textbf{Open Area}} \\
        \cmidrule(lr){2-7} \cmidrule(lr){8-13} \cmidrule(lr){14-19}
        \textbf{Method}
        & \textbf{MAE}$\downarrow$ & \textbf{CEP}$\downarrow$ & \textbf{R95}$\downarrow$ & \textbf{KS}$\downarrow$ & \textbf{TS}$\uparrow$ & \textbf{CT}$\uparrow$
        & \textbf{MAE}$\downarrow$ & \textbf{CEP}$\downarrow$ & \textbf{R95}$\downarrow$ & \textbf{KS}$\downarrow$ & \textbf{TS}$\uparrow$ & \textbf{CT}$\uparrow$
        & \textbf{MAE}$\downarrow$ & \textbf{CEP}$\downarrow$ & \textbf{R95}$\downarrow$ & \textbf{KS}$\downarrow$ & \textbf{TS}$\uparrow$ & \textbf{CT}$\uparrow$ \\
        \midrule
        TDoA
        & 7.019 & 6.283 & 15.615 & --- & --- & ---
        & 2.443 & 1.867 & 6.122 & --- & --- & ---
        & 6.276 & 6.242 & 10.923 & --- & --- & --- \\
        \midrule
        FP
        & 0.520 & 0.389 & 1.312 & --- & --- & ---
        & 0.590 & 0.497 & 1.342 & --- & --- & ---
        & 0.739 & 0.607 & 1.808 & --- & --- & --- \\
        \midrule
        PCA
        & 3.870 & 3.759 & 7.172 & 0.292 & 0.928 & 0.960
        & 2.750 & 2.289 & 5.567 & 0.313 & 0.919 & 0.956
        & 2.203 & 2.014 & 4.362 & 0.339 & 0.872 & 0.920 \\
        Isomap*
        & 4.455 & 4.143 & 9.095 & 0.405 & 0.962 & 0.956
        & 1.896 & 1.709 & 3.958 & 0.212 & 0.951 & 0.949
        & 2.016 & 1.867 & 4.006 & 0.321 & 0.912 & 0.909 \\
        t-SNE*
        & 3.461 & 3.175 & 7.511 & 0.295 & \textbf{0.996} & 0.977
        & 3.733 & 2.887 & 9.530 & 0.399 & \textbf{0.995} & 0.966
        & 1.665 & 1.336 & 4.164 & 0.282 & \textbf{0.995} & 0.937 \\
        UMAP*
        & 2.513 & \textbf{1.959} & 5.981& 0.195 & 0.993 & 0.983
        & 3.118 & 2.534 & 6.917 & 0.328 & 0.989 & 0.958
        & 1.798 & 1.449 & 4.336 & 0.279 & 0.989 & 0.937 \\
        AE
        & 4.443 & 4.383 & 8.663 & 0.366 & 0.985 & 0.974
        & 1.894 & 1.683 & 3.934 & 0.230 & 0.957 & 0.959
        & 3.177 & 2.926 & 6.285 & 0.461 & 0.927 & 0.914 \\
        ConstAE
        & 5.236 & 4.727 & 11.861 & 0.419 & 0.982 & 0.960
        & 2.086 & 1.784 & 5.017 & 0.474 & 0.944 & 0.931
        & 2.556 & 2.313 & 5.521 & 0.425 & 0.843 & 0.916 \\
        UAGeo
        & 3.451 & 2.631 & 9.317 & 0.345 & 0.957 & 0.962
        & 1.422 & 1.412 & 2.486 & 0.139 & 0.931 & 0.953
        & 1.440 & 1.322 & 2.899 & 0.246 & 0.918 & 0.935 \\
        PSSN
        & 3.072 & 2.827 & 6.046 & 0.248 & 0.951 & 0.971
        & 1.974 & 1.837 & 3.725 & 0.277 & 0.920 & 0.961
        & 2.545 & 2.470 & 4.400 & 0.303 & 0.857 & 0.901 \\
        Siamese
        & 2.931 & 2.688 & 5.960 & 0.244 & 0.959 & 0.970
        & 2.430 & 2.158 & 5.133 & 0.344 & 0.921 & 0.938
        & 1.725 & 1.604 & 3.463 & 0.311 & 0.880 & 0.921 \\
        Siamese-G
        & 4.599 & 4.506 & 8.373 & 0.421 & 0.973 & 0.976
        & 1.697 & 1.667 & 3.058 & 0.199 & 0.940 & 0.957
        & 1.657 & 1.526 & 3.325 & 0.266 & 0.907 & 0.935 \\
        Triplet
        & \textbf{2.440} & 2.028 & \textbf{5.118} & \textbf{0.171} & 0.990 & \textbf{0.992}
        & \textbf{1.173} & \textbf{1.166} & \textbf{1.878} & \textbf{0.111} & 0.971 & \textbf{0.967}
        & \textbf{1.182} & \textbf{1.128} & \textbf{2.217} & \textbf{0.190} & 0.971 & \textbf{0.966} \\
    \end{tabular}
    }
    \label{tab:metrics-all}
    \vspace{-0.5cm}
\end{table*}

\section{Benchmark Evaluation}\label{sec:results}

We evaluate \dsname{} as a benchmark for geometric, supervised, unsupervised, and self-supervised 5G indoor localization.
We not only compare localization accuracy, but also analyze how method classes perform under severe multipath.

\subsection{Experimental Protocol and Methods}

For all learning-based methods, we use the magnitude of the \acp{cir} as input.
We apply temporal smoothing to the \acp{cir} and median filtering to the \ac{toa} estimates using a 100\,ms window.
The \acp{cir} are then aligned based on the \ac{toa} estimates and normalized to unit norm per sample.
This reduces the influence of autonomous \ac{ue} transmit-power control while preserving the relative multipath structure.
We evaluate 3 dataset splits.
The \textit{full} split uses takes \#1--\#6 for training and take \#7 for testing.
The \textit{corridor} split uses take \#4 for training and take \#1 for testing, focusing on the lower metallic corridor.
The \textit{open} split uses take \#5 for training and take \#2 for testing, focusing on the open area.
Pretraining methods are evaluated only on the full split.
For UAGeo on the full split, we use take~\#8 for training as its acceleration loss requires a continuous trajectory.
Methods that rely on pairwise distance matrices, namely UAGeo and the Siamese variants, are trained on data downsampled by a factor of ten due to their $O(n^2)$ memory requirements.
We compare four groups of methods.
First, we evaluate geometric \ac{tdoa} using a single-shot Levenberg--Marquardt solver~\cite{ls-tdoa}.
Second, we train a supervised fingerprinting model, which serves as an upper bound as it uses reference positions during training.
Third, we evaluate unsupervised channel charting and dimensionality-reduction methods, including PCA~\cite{pca}, t-SNE~\cite{tsne}, Isomap~\cite{isomap}, UMAP~\cite{umap}, AE~\cite{cc}, ConstAE~\cite{ae-cc}, PSSN~\cite{zhao_signature_2024}, UAGeo~\cite{euchner2024uncertainty}, Siamese~\cite{siamese-cc}, Siamese-G~\cite{stahlke2023indoor}, and Triplet networks~\cite{triplet-cc}.
Finally, we evaluate self-supervised wireless pretraining methods, namely WiMAE, ContraWiMAE~\cite{wimae}, LWM~\cite{lwm}, and SpaRTran~\cite{sparstran}.
For supervised localization and \ac{tdoa}, we directly report the localization error.
For channel charting methods, the learned charts are aligned to the reference coordinates using an optimal affine transform before computing localization metrics.
We report the \ac{mae}, circular error probable (CEP), 95th percentile error (R95), Kruskal stress (KS), trustworthiness (TS), and continuity (CT).
Non-parametric methods marked with * are fitted jointly on train and test samples and should therefore be interpreted as transductive baselines.

\subsection{Localization Performance}

Tab.~\ref{tab:metrics-all} summarizes the localization results on all three dataset splits.
The results show that \dsname{} is a challenging benchmark for current indoor localization methods.
On the full split, geometric \ac{tdoa} reaches 7.02\,m \ac{mae}, confirming that direct geometric localization is strongly affected by \ac{nlos} propagation and corrupted \ac{toa} estimates.
In contrast, supervised fingerprinting achieves 0.52\,m \ac{mae}, showing that the \acp{cir} contain strong spatial information when sufficient labels are available.
Unsupervised methods remain substantially less accurate than supervised fingerprinting.
On the full split, the best unsupervised method is the Triplet network with 2.44\,m \ac{mae}.
This is a clear improvement over simple dimensionality-reduction and autoencoder baselines, but still leaves a large gap to supervised fingerprinting.
The gap indicates that the main difficulty is not the absence of spatial information in the radio measurements, but the inability of current unsupervised objectives to recover this information reliably under severe industrial multipath.
The subarea splits reveal that the propagation regime strongly affects the ranking of methods.
In the corridor split, the Triplet network performs best with 1.17\,m \ac{mae}.
We attribute this to its temporal-proximity supervision, which provides robust local constraints even when \ac{cir}-based dissimilarities are distorted by abrupt multipath changes.
In the open split, several neighborhood-based and geodesic methods perform competitively, with Siamese-G, UAGeo, t-SNE, and UMAP reaching around 1.4--1.8\,m \ac{mae}.
This suggests that current channel-charting assumptions are more appropriate in less obstructed regions, but become fragile in the metallic corridors and in the full heterogeneous environment.

\subsection{Failure Modes of Channel Charting}


\begin{figure}[b]
    \centering
    \vspace{-0.6cm}
    \includegraphics[trim=0 0 0 0, clip, width=0.9\linewidth]{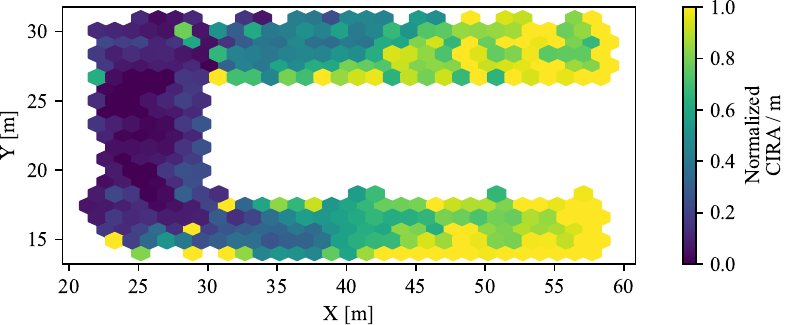}
    \vspace{-0.35cm}
    \caption{Norm. CIRA metric change per real-world distance. Larger changes in the corridors indicate stronger spatial variability of the radio channel.}
    \label{fig:cira_vs_distance}
\end{figure}

\begin{figure*}[t]
    \centering
    \subfloat[Ground Truth]{\includegraphics[width=0.21\linewidth]{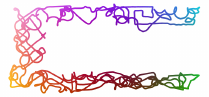}}\hfill
    \subfloat[PCA]{\includegraphics[width=0.21\linewidth]{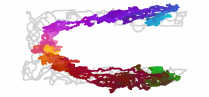}}\hfill
    \subfloat[Isomap]{\includegraphics[width=0.21\linewidth]{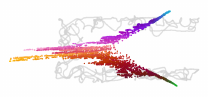}}\label{isomap}\hfill
    \subfloat[t-SNE]{\includegraphics[width=0.21\linewidth]{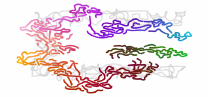}}\\[-8pt]

    \subfloat[UMAP]{\includegraphics[width=0.21\linewidth]{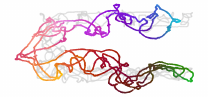}}\hfill
    \subfloat[AE]{\includegraphics[width=0.21\linewidth]{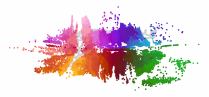}}\hfill
    \subfloat[ConstAE]{\includegraphics[width=0.21\linewidth]{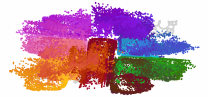}}\hfill
    \subfloat[UAGeo]{\includegraphics[width=0.21\linewidth]{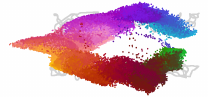}}\\[-8pt]

    \subfloat[PSSN]{\includegraphics[width=0.21\linewidth]{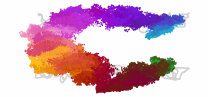}}\hfill
    \subfloat[Siamese]{\includegraphics[width=0.21\linewidth]{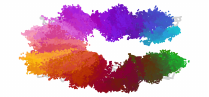}}\hfill
    \subfloat[Siamese-G]{\includegraphics[width=0.21\linewidth]{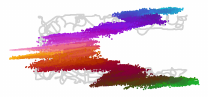}}\hfill
    \subfloat[Triplet]{\includegraphics[width=0.21\linewidth]{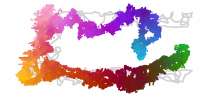}}
    \caption{Channel charts colorized by reference position after optimal affine alignment.}
    \label{fig:cc-results}
    \vspace{-0.5cm}
\end{figure*}

Channel charting methods assume that radio dissimilarities vary smoothly with physical distance.
This assumption is weakened in \dsname{}.
As shown in Fig.~\ref{fig:cira_vs_distance}, the CIRA metric changes more rapidly per meter in the metallic corridors than in the open area.
This indicates that physically nearby positions produce substantially different \acp{cir} when dominant reflections appear or disappear over short distances.
The learned channel charts in Fig.~\ref{fig:cc-results} illustrate the resulting failure modes.
Several methods distort, stretch, or partially separate corridor regions, showing that metal-induced multipath does not merely add noise but changes the structure of the learned radio manifold.
Geodesic methods additionally struggle with the non-convex topology of the recording area, as the C-shaped geometry encourages embeddings that unwrap the environment rather than preserve its physical layout.
This can be especially seen at the Siamese-G and Isomap channel charts in Fig.~\ref{fig:cc-results} (the affine transformation stretches the channel charts along the x-axis, which partially compensates for this effect visually).
These effects explain why high local structure metrics such as trustworthiness or continuity do not always translate into low localization error after affine alignment.

\subsection{Self-Supervised Pretraining}

\begin{figure}[b]
    \centering
    \vspace{-0.4cm}
    \includegraphics[trim=0 0 0 0, width=\linewidth]{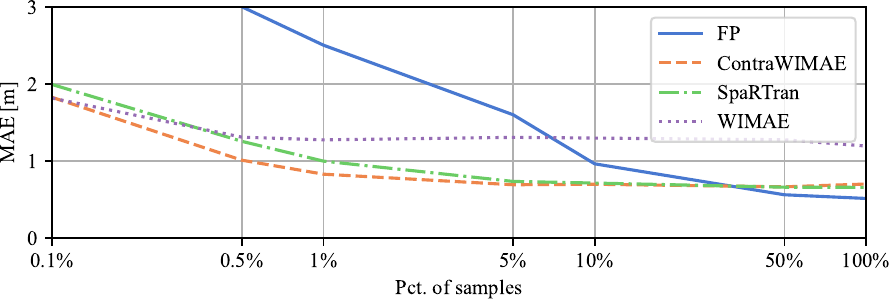}
    \vspace{-0.8cm}
    \caption{Localization performance after self-supervised pretraining and supervised fine-tuning with different fractions of labeled samples.}
    \label{fig:pretrain}
\end{figure}

The pretraining results in Fig.~\ref{fig:pretrain} show that self-supervised learning substantially reduces the amount of labeled data required for fingerprinting.
With only 1\% of the labeled training samples, ContraWiMAE reaches 0.83\,m \ac{mae}, outperforming the supervised fingerprinting baseline trained with 10\% of the labels, which reaches 0.97\,m.
SpaRTran shows a similar trend and reaches 1.00\,m \ac{mae} with 1\% labels.
This corresponds to an approximate tenfold reduction in labeling effort.
At 100\% labels, SpaRTran and ContraWiMAE reach 0.66\,m and 0.67\,m \ac{mae}, respectively, approaching but not matching the supervised fingerprinting baseline at 0.52\,m.
This suggests that pretraining learns useful spatial radio representations, but still discards or fails to exploit some fine-grained localization information contained in the \acp{cir}.
WiMAE provides more modest gains and saturates at higher error levels, while LWM collapsed during fine-tuning in our experiments.
Overall, self-supervised pretraining offers a practical compromise between label efficiency and accuracy, but does not fully solve robust localization under severe industrial multipath.

\subsection{Discussion}
The benchmark results support three main conclusions.
First, \dsname{} is substantially more challenging than clean or mildly obstructed localization settings: \ac{tdoa} performs poorly on the full dataset, and unsupervised methods remain far from supervised fingerprinting.
Second, the strong performance of supervised fingerprinting shows that the \acp{cir} contain rich spatial information, but current unsupervised objectives struggle to recover it under irregular metal-induced multipath.
Third, self-supervised pretraining is a promising direction for reducing deployment cost, but the remaining gap to full supervision highlights the need for more robust representation learning methods for industrial 5G localization.

\section{Conclusion} \label{sec:conclusion}

We presented \dsname{}, a public real-world 5G NR \ac{cir} benchmark for AI-based indoor localization in an operational railway maintenance facility.
The dataset covers 811\,m$^2$ with eight distributed \acp{ru}, robot-mounted \ac{ue} trajectories, high-accuracy iGPS reference positions, and severe metal-induced multipath caused by train compartments and maintenance infrastructure.
The benchmark confirms that \dsname{} is challenging for current localization methods.
On the full split, geometric \ac{tdoa} reaches 7.02\,m \ac{mae}, the best unsupervised method reaches 2.44\,m, and supervised fingerprinting achieves 0.52\,m.
This gap shows that the \acp{cir} contain strong spatial information, but that current unsupervised methods struggle to exploit it under severe industrial multipath.
Our analysis further shows that metallic train compartments distort the radio manifold by causing abrupt spatial changes in the \acp{cir}.
Self-supervised pretraining reduces labeling requirements by up to one order of magnitude, but does not fully close the gap to supervised fingerprinting.
By releasing \dsname{}, we provide a challenging benchmark for developing robust AI-based 5G indoor localization methods for real industrial environments.

\AtNextBibliography{\small}
\printbibliography

@article{Laoudias2018,
  author  = {Laoudias, Christos and Moreira, Adriano J. C. and Kim, Sunwoo and Lee, Sang-Woo and Wirola, Lauri and Fischione, Carlo},
  title   = {{A Survey of Enabling Technologies for Network Localization, Tracking, and Navigation}},
  journal = {IEEE Commun. Surveys Tuts.},
  year    = {2018},
  volume  = {20},
  number  = {4},
  pages   = {3607--3644},
  doi     = {10.1109/COMST.2018.2855063}
}

@article{Dardari2015,
  author  = {Dardari, Davide and Closas, Pau and Djuri{\'c}, Petar M.},
  title   = {{Indoor Tracking: Theory, Methods, and Technologies}},
  journal = {IEEE Trans. Veh. Technol.},
  year    = {2015},
  volume  = {64},
  number  = {4},
  pages   = {1263--1278},
  doi     = {10.1109/TVT.2015.2403868}
}

@article{cc,
  author  = {Studer, Christoph and Medjkouh, Sa{\"\i}d and G{\"o}n{\"u}lta{\c{s}}, Emre and Goldstein, Tom and Tirkkonen, Olav},
  title   = {{Channel Charting: Locating Users Within the Radio Environment Using Channel State Information}},
  journal = {IEEE Access},
  year    = {2018},
  volume  = {6},
  number  = {},
  pages   = {47682--47698},
  doi     = {10.1109/ACCESS.2018.2866979}
}

@article{pdr-cc,
  author  = {Stahlke, Maximilian and Yammine, George and Feigl, Tobias and Eskofier, Bjoern M. and Mutschler, Christopher},
  title   = {{Velocity-Based Channel Charting With Spatial Distribution Map Matching}},
  journal = {IEEE J. Indoor Seamless Position. Navig.},
  year    = {2024},
  volume  = {2},
  number  = {},
  pages   = {230--239},
  doi     = {10.1109/JISPIN.2024.3424768}
}

@inproceedings{triplet-cc,
  author    = {Ferrand, Paul and Decurninge, Alexis and Ordo{\~n}ez, Luis G. and Guillaud, Maxime},
  title     = {{Triplet-Based Wireless Channel Charting}},
  booktitle = {Proc. IEEE Glob. Commun. Conf. (GLOBECOM)},
  year      = {2020},
  address   = {Taipei, Taiwan},
  pages     = {1--6},
  doi       = {10.1109/GLOBECOM42002.2020.9322643}
}

@article{adp-metric,
  author  = {Stephan, Phillip and Euchner, Florian and ten Brink, Stephan},
  title   = {{Angle-Delay Profile-Based and Timestamp-Aided Dissimilarity Metrics for Channel Charting}},
  journal = {IEEE Trans. Commun.},
  year    = {2024},
  volume  = {72},
  number  = {9},
  pages   = {5611--5625},
  doi     = {10.1109/TCOMM.2024.3386571}
}

@misc{alkhateeb2019deepmimo,
  author        = {Alkhateeb, Ahmed},
  title         = {{DeepMIMO: A Generic Deep Learning Dataset for Millimeter Wave and Massive MIMO Applications}},
  year          = {2019},
  eprint        = {1902.06435},
  archivePrefix = {arXiv},
  primaryClass  = {eess.SP},
  doi           = {10.48550/arXiv.1902.06435}
}

@inproceedings{makhlouf2025multi,
  author    = {Makhlouf, Mohamed El Mehdi and Guillaud, Maxime and Vindas, Yamil},
  title     = {{Multi-Cell Outdoor Channel State Information Dataset (MOCSID)}},
  booktitle = {Proc. Joint Eur. Conf. Netw. Commun. \& 6G Summit (EuCNC/6G Summit)},
  year      = {2025},
  address   = {Pozna{\'n}, Poland},
  pages     = {85--90},
  doi       = {10.5281/zenodo.15294869}
}

@inproceedings{dichasus2021,
  author    = {Euchner, Florian and Gauger, Marc and D{\"o}rner, Sebastian and ten Brink, Stephan},
  title     = {{A Distributed Massive MIMO Channel Sounder for ``Big CSI Data''-Driven Machine Learning}},
  booktitle = {Proc. Int. ITG WSA},
  year      = {2021},
  address   = {French Riviera, France},
  pages     = {289--294},
  doi       = {10.48550/arXiv.2206.15302}
}

@inproceedings{espargos2024,
  author    = {Euchner, Florian and ten Brink, Stephan},
  title     = {{ESPARGOS: Phase-Coherent WiFi CSI Datasets for Wireless Sensing Research}},
  booktitle = {Proc. Kleinheubach Conf.},
  year      = {2024},
  address   = {Miltenberg, Germany},
  pages     = {1--4},
  doi       = {10.23919/IEEECONF64570.2024.10739065}
}

@inproceedings{caez-5g2025,
  author    = {Wiesmayr, Reinhard and Zumegen, Frederik and Taner, Sueda and Dick, Chris and Studer, Christoph},
  title     = {{CSI-Based User Positioning, Channel Charting, and Device Classification With an NVIDIA 5G Testbed}},
  booktitle = {Proc. Asilomar Conf. Signals Syst. Comput.},
  year      = {2025},
  address   = {Pacific Grove, CA},
  pages     = {1226--1232},
  doi       = {10.1109/IEEECONF67917.2025.11443744}
}

@inproceedings{caez-wifi2024,
  author    = {Zumegen, Frederik and Studer, Christoph},
  title     = {{A Software-Defined and Distributed Wi-Fi Channel-State Information Acquisition Testbed}},
  booktitle = {Proc. Asilomar Conf. Signals Syst. Comput.},
  year      = {2024},
  address   = {Pacific Grove, CA},
  pages     = {338--344},
  doi       = {10.1109/IEEECONF60004.2024.10942693}
}

@article{pca,
  author  = {Ma{\'c}kiewicz, Andrzej and Ratajczak, Waldemar},
  title   = {{Principal Components Analysis (PCA)}},
  journal = {Comput. Geosci.},
  year    = {1993},
  volume  = {19},
  number  = {3},
  pages   = {303--342},
  doi     = {10.1016/0098-3004(93)90090-R}
}

@article{tsne,
  author  = {Laurens van der Maaten and Geoffrey Hinton},
  title   = {Visualizing Data using t-SNE},
  journal = {Journal of Machine Learning Research},
  year    = {2008},
  volume  = {9},
  number  = {86},
  pages   = {2579--2605},
  url     = {http://jmlr.org/papers/v9/vandermaaten08a.html}
}

@misc{umap,
  author        = {McInnes, Leland and Healy, John and Melville, James},
  title         = {{UMAP: Uniform Manifold Approximation and Projection for Dimension Reduction}},
  year          = {2018},
  eprint        = {1802.03426},
  archivePrefix = {arXiv},
  primaryClass  = {stat.ML},
  doi           = {10.48550/arXiv.1802.03426}
}

@article{isomap,
  author  = {Tenenbaum, Joshua B. and de Silva, Vin and Langford, John C.},
  title   = {{A Global Geometric Framework for Nonlinear Dimensionality Reduction}},
  journal = {Science},
  year    = {2000},
  volume  = {290},
  number  = {5500},
  pages   = {2319--2323},
  doi     = {10.1126/science.290.5500.2319}
}

@inproceedings{ls-tdoa,
  author    = {Mensing, Christian and Plass, Simon},
  title     = {{Positioning Algorithms for Cellular Networks Using TDoA}},
  booktitle = {Proc. IEEE ICASSP},
  year      = {2006},
  address   = {Toulouse, France},
  pages     = {4-18},
  doi       = {10.1109/ICASSP.2006.1661018}
}

@inproceedings{euchner2024uncertainty,
  author    = {Euchner, Florian and Stephan, Phillip and ten Brink, Stephan},
  title     = {{Uncertainty-Aware Dimensionality Reduction for Channel Charting With Geodesic Loss}},
  booktitle = {Proc. Asilomar Conf. Signals Syst. Comput.},
  year      = {2024},
  address   = {Pacific Grove, CA},
  pages     = {1665--1672},
  doi       = {10.1109/IEEECONF60004.2024.10942819}
}

@article{stahlke2023indoor,
  author  = {Stahlke, Maximilian and Yammine, George and Feigl, Tobias and Eskofier, Bjoern M. and Mutschler, Christopher},
  title   = {{Indoor Localization With Robust Global Channel Charting: A Time-Distance-Based Approach}},
  journal = {IEEE Trans. Mach. Learn. Commun. Netw.},
  year    = {2023},
  volume  = {1},
  number  = {},
  pages   = {3--17},
  doi     = {10.1109/TMLCN.2023.3256964}
}

@misc{fhg5g,
  author = {{Fraunhofer Institute for Integrated Circuits IIS}},
  title  = {{Fingerprinting Dataset for Positioning}},
  year   = {2024},
  url          = {https://www.iis.fraunhofer.de/en/ff/lv/dataanalytics/pos/fingerprinting-dataset-for-positioning.html},
  note   = {Dataset}
}

@inproceedings{siamese-cc,
  author    = {Lei, Eric and Casta{\~n}eda, Oscar and Tirkkonen, Olav and Goldstein, Tom and Studer, Christoph},
  title     = {{Siamese Neural Networks for Wireless Positioning and Channel Charting}},
  booktitle = {Proc. Annu. Allerton Conf. Commun. Control Comput.},
  year      = {2019},
  address   = {Monticello, IL},
  pages     = {200--207},
  doi       = {10.1109/ALLERTON.2019.8919897}
}

@article{mamimo,
  author  = {Li, Chenglong and De Bast, Sibren and Tanghe, Emmeric and Pollin, Sofie and Joseph, Wout},
  title   = {{Toward Fine-Grained Indoor Localization Based on Massive MIMO-OFDM System: Experiment and Analysis}},
  journal = {IEEE Sensors J.},
  year    = {2022},
  volume  = {22},
  number  = {6},
  pages   = {5318--5328},
  doi     = {10.1109/JSEN.2021.3111986}
}

@article{conti-3gpp-compliant-2024,
  author  = {Conti, Andrea and Torsoli, Gianluca and G{\'o}mez-Vega, Carlos A. and Vaccari, Alessandro and Mazzini, Gianluca and Win, Moe Z.},
  title   = {{3GPP-Compliant Datasets for xG Location-Aware Networks}},
  journal = {IEEE Open J. Veh. Technol.},
  year    = {2024},
  volume  = {5},
  number  = {},
  pages   = {473--484},
  doi     = {10.1109/OJVT.2023.3340993}
}

@inproceedings{zhao_signature_2024,
  author    = {Zhao, Longhai and Yang, Yunchuan and Xiong, Qi and Wang, He and Yu, Bin and Sun, Feifei and Sun, Chengjun},
  title     = {{A Signature Based Approach Towards Global Channel Charting With Ultra Low Complexity}},
  booktitle = {Proc. IEEE ICC Workshops},
  year      = {2024},
  address   = {Denver, CO},
  pages     = {667--672},
  doi       = {10.1109/ICCWorkshops59551.2024.10615753}
}

@misc{wimae,
  author        = {Guler, Berkay and Geraci, Giovanni and Jafarkhani, Hamid},
  title         = {{A Multi-Task Foundation Model for Wireless Channel Representation Using Contrastive and Masked Autoencoder Learning}},
  year          = {2025},
  eprint        = {2505.09160},
  archivePrefix = {arXiv},
  primaryClass  = {eess.SP},
  doi           = {10.48550/arXiv.2505.09160}
}

@misc{lwm,
  author        = {Alikhani, Sadjad and Charan, Gouranga and Alkhateeb, Ahmed},
  title         = {{Large Wireless Model (LWM): A Foundation Model for Wireless Channels}},
  year          = {2024},
  eprint        = {2411.08872},
  archivePrefix = {arXiv},
  primaryClass  = {eess.SP},
  doi           = {10.48550/arXiv.2411.08872}
}

@misc{sparstran,
  author        = {Ott, Jonathan and Stahlke, Maximilian and Feigl, Tobias and Eskofier, Bjoern M. and Mutschler, Christopher},
  title         = {{Simplicity Is Key: An Unsupervised Pretraining Approach for Sparse Radio Channels}},
  year          = {2025},
  eprint        = {2505.13055},
  archivePrefix = {arXiv},
  primaryClass  = {cs.LG},
  doi           = {10.48550/arXiv.2505.13055}
}

@inproceedings{ae-cc,
  author    = {Huang, Pengzhi and Casta{\~n}eda, Oscar and G{\"o}n{\"u}lta{\c{s}}, Emre and Medjkouh, Sa{\"\i}d and Tirkkonen, Olav and Goldstein, Tom and Studer, Christoph},
  title     = {{Improving Channel Charting With Representation-Constrained Autoencoders}},
  booktitle = {Proc. IEEE Int. Workshop SPAWC},
  year      = {2019},
  address   = {Cannes, France},
  pages     = {1--5},
  doi       = {10.1109/SPAWC.2019.8815478}
}

@misc{gassner2021opencsi,
  author        = {Gassner, Arthur and Musat, Claudiu and Rusu, Alexandru and Burg, Andreas},
  title         = {{OpenCSI: An Open-Source Dataset for Indoor Localization Using CSI-Based Fingerprinting}},
  year          = {2021},
  eprint        = {2104.07963},
  archivePrefix = {arXiv},
  primaryClass  = {cs.NI},
  doi           = {10.48550/arXiv.2104.07963}
}

@inproceedings{debast2020mamimo,
  author    = {De Bast, Sibren and Pollin, Sofie},
  title     = {{MaMIMO CSI-Based Positioning Using CNNs: Peeking Inside the Black Box}},
  booktitle = {Proc. IEEE ICC Workshops},
  year      = {2020},
  address   = {Dublin, Ireland},
  pages     = {1--6},
  doi       = {10.1109/ICCWorkshops49005.2020.9145412}
}

@misc{debast2021ultradense,
  author    = {De Bast, Sibren and Pollin, Sofie},
  title     = {{Ultra Dense Indoor MaMIMO CSI Dataset}},
  year      = {2021},
  publisher = {IEEE DataPort},
  doi       = {10.21227/nr6k-8r78},
  note      = {Dataset}
}

@misc{hofmann2009device,
  title={Device and method for determining a time of arrival of a receive sequence},
  author={Hofmann, Guenter and Breiling, Marco},
  year={2009},
  month=dec # "~1",
  publisher={Google Patents},
  note={US Patent 7,627,063}
}

\end{document}